\documentclass[prx,aps,twocolumn,superscriptaddress,letterpaper,longbibliography]{revtex4-1}
\usepackage{amssymb}
\usepackage{mathtools}
\usepackage[pdftex]{graphicx}
\usepackage[usenames, dvipsnames, svgnames, table]{xcolor}
\usepackage[colorlinks=true, citecolor=Blue,
            linkcolor=BrickRed, urlcolor=ForestGreen]{hyperref}
\usepackage{txfonts}
\usepackage{mathrsfs}
\usepackage{bm}
\begin{document}

%%%%%% Some definition %%%%%%

\newcommand{\be}{\begin{equation}}
\newcommand{\ee}{\end{equation}}
\newcommand{\R}[1]{\textcolor{red}{#1}}
\newcommand{\B}[1]{\textcolor{blue}{#1}}
\newcommand{\fixme}[1]{\textcolor{orange}{#1}}

%% losses
\def\earm{\epsilon_{\rm arm}}
\def\esrc{\epsilon_{\rm src}}
\def\eext{\epsilon_{\rm ext}}
\def\eint{\epsilon_{\rm int}}
\def\eout{\epsilon_{\rm out}}

%% transmissions
\def\Titm{T_{\rm itm}}
\def\Tsrc{T_{\rm src}}

%% matrices
\def\Mrot{{\bf M}_{\rm rot}}
\def\Msqz{{\bf M}_{\rm sqz}}
\def\Mopt{{\bf M}_{\rm opt}}
\def\Mio{{\bf M}_{\rm io}}
\def\Mc{{\bf M}_{\rm c}}

%%%%%% Title %%%%%%

\title{Quantum limit for laser interferometric gravitational wave detectors from optical
dissipation}

\author{Haixing Miao}
\affiliation{School of Physics and Astronomy,
University of Birmingham, Birmingham, B15 2TT, United Kingdom}
\author{Nicolas D. Smith}
\affiliation{California Institute of Technology, Pasadena, CA 91125, USA}
\author{Matthew Evans}
\affiliation{Massachusetts Institute of Technology, Cambridge, MA 02139, USA}

%%%%%% Abstract %%%%%%

\begin{abstract}
We derive a quantum limit to the sensitivity of 
laser interferometric gravitational-wave detectors 
from optical-loss-induced dissipation, 
analogous to the sensitivity limit from the mechanical dissipation. 
It applies universally to different interferometer configurations, 
and cannot be surpassed unless optical property is improved.  
This
result provides an answer to the long-standing question of how far we can push the detector sensitivity given the
state-of-the-art optics.
\end{abstract}

\maketitle

%%%%%% Introduction %%%%%%

{\it Introduction --- }Advanced gravitational-wave (GW) detectors are 
long-baseline interferometers with suspended mirrors which act as test masses
for probing spacetime dynamics. Quantum noise, arising from quantum fluctuations of the optical field,
 is one of the sources of noise that limits the sensitivity of such instruments.
In particular, the phase fluctuation gives rise to the shot noise, while the amplitude
fluctuation exerts a random force on the test masses and
induces the quantum radiation pressure noise. 
These two types of quantum noise, when uncorrelated,
lead to the Standard Quantum Limit (SQL)\,\cite{SQL1, Braginsky92} 
of which the power spectral density is 
\begin{equation}
S_{hh}^{\rm SQL}(\Omega)=\frac{8\hbar}{M \Omega^2L^2}\,,
\end{equation}
where $\Omega$ is the angular frequency of the GW
signal, 
$M$ is the mass of the test-mass mirror, and $L$ is the interferometer arm length.

\begin{figure}[!b]
  \includegraphics[width=0.7\columnwidth]{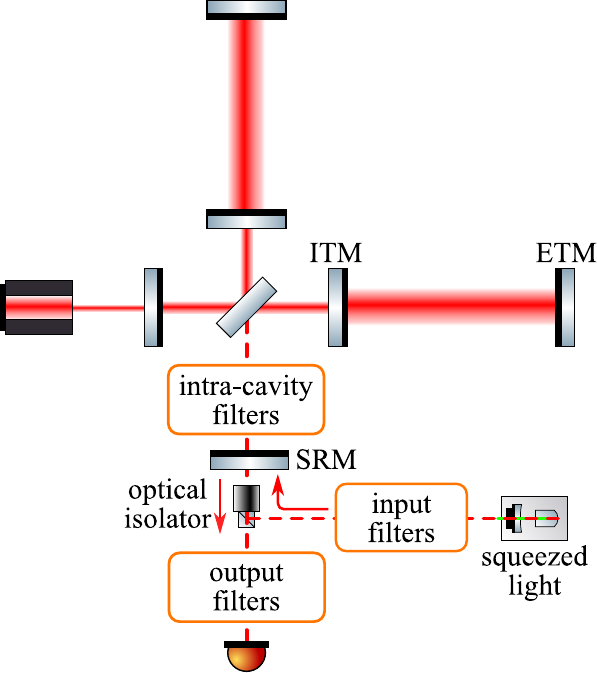}
    \caption{Illustration of the general QND scheme
    with optical filters added to the typical configuration 
     of advanced GW detectors---a dual-recycled
    Michelson interferometer. ITM: input test mass, ETM: end test 
    mass, and SRM: signal-recycling mirror.}
\label{fig:config}
\end{figure}

Despite its name, the SQL is not a true limit:
 it can be surpassed with a wide class of so-called
 quantum-non-demolition (QND) schemes 
 (cf. review articles\,\cite{Braginsky80, lrr-2012-5, Miao2014}). 
These techniques usually involve modifications to the optical configuration 
of current-generation GW detectors
by introducing extra optical filters. These filters can be a cascade of both passive 
 Fabry-Perot cavities and active cavities which have external energy input\,\cite{Wicht1997, Miao2015a, Zhou2015}. 
For example, together with a squeezed light source,
 a passive filter cavity can be used for producing
 the frequency-dependent squeezing\,\cite{Kimble01, Oelker2016, Schnabel2017}. 
The general scheme is illustrated in Fig.\,\ref{fig:config}.

The Quantum Cram\'er-Rao Bound (QCRB)\,\cite{Helstrom1967, 
Holevo2011} is
a sensitivity limit which, unlike the SQL, is inviolable.
In the context of GW detection with laser interferometers,
the QCRB is also called the energetic or 
fundamental quantum limit\,\cite{Braginsky2000EQL, Tsang2011}.
The spectral representation of the QCRB is 
\begin{equation}\label{eq:FQL}
S_{hh}^{\rm QCRB}(\Omega) =
\frac{\hbar^2 c^2}{2S_{PP}(\Omega)L^2 }\,,
\end{equation}
in which $S_{PP}$ is the spectral density of the power
fluctuations in the interferometer arms.
As shown in Ref.\,\cite{Miao2017}, the QCRB can be approached
 in a lossless systems with optimal frequency-dependent homodyne readout, 
which can be realized with proper output filters\,\cite{Kimble01}. This 
result has been generalised to laser interferometers with multiple carrier 
frequencies\,\cite{Branford2018}. 

According to the QCRB, the sensitivity is ultimately bounded by the 
power or equivalently energy fluctuation inside the arm cavities. This can be
intuitively understood from the fact that we want to measure
the phase or timing difference between the two interferometer arms 
accurately, and a large uncertainty in
the photon number or energy is needed, due to the
number-phase or energy-time uncertainty relation.
Note, however, that increasing $S_{PP}$ is only advantageous if
 a minimum uncertainty state, or at least nearly minimum uncertainty state, can be maintained.
Minimum uncertainty states are, however, very delicate and easily destroyed by
 optical losses which lead to decoherence.

The power fluctuation $S_{PP}$ can be enhanced,
 and thus the QCRB be reduced, with a variety of approaches.
The most obvious approach is to increase the optical power in the interferometer,
 but this is limited by thermal lensing\,\cite{Lawrence02}, 
 alignment stability\,\cite{Hirose:10}, and parametric instabilities\,\cite{Braginsky2001, Evans15}. 
A second approach is to introduce squeezed states of light 
generated by nonlinear optical processes\,\cite{Schnabel2017}
at the readout port (cf. Fig.\,\ref{fig:config}).
In addition to squeezing the vacuum fluctuations which-- enter the interferometer,
 the radiation  pressure coupling between the optical field and the test masses 
produces squeezing internall.y
This process, known as ``pondermotive squeezing''\,\cite{Kimble01, 
Aspelmeyer2014},
 is actually the cause of the radiation pressure noise. We can it turn into a resource 
 for enhancing the detector sensitivity by detuning the signal-recycling cavity (SRC)
 formed by ITM and SRM away from the perfect
 resonance. This is the so-called optical spring effect which was studied extensively by focusing on the dynamics of the test masses\,\cite{BuCh2002}. In the Appendix \ref{app:OS-sqz}, we illustrate an equivalent picture mentioned in Ref.\,\cite{Miao2017}
from the perspective of the internal ponderomotive squeezing. 

In principle, we can combine the above-mentioned approaches to 
increase $S_{PP}$ and make the QCRB vanishingly small,
 which implies an unbounded sensitivity.
A clear example of this would be the injection of a very strongly squeezed
 state which could increase $S_{PP}$ by more than an order of magnitude.
In the presence of optical losses, a highly squeezed state will cease to be
 a minimum uncertainty state due to decoherence, and the QCRB in the lossless case 
 will not be a relevant bound.

In this paper, we present a new general limit to GW detectors which
 cannot be surpassed and is more constraining than the QCRB for 
 realistic interferometer configurations with optical losses.
As we will show, the optical losses lead to the following sensitivity limit,
 to the first order of the loss 
parameter $\epsilon$:
\begin{equation}\label{eq:Shh_loss}
S_{hh}^{\epsilon}= \frac{\hbar\,  c^2 }{4 L^2 \omega_0
   P}\left[ \earm+\left(1+\frac{\Omega^2}{\gamma^2}\right)\frac{\Titm \esrc}{4}+\alpha\, 
    \Tsrc\eext\right]\,. 
\end{equation}
Here $P$ is the optical power inside each arm (assumed to be equal); 
$\omega_0$ is the laser frequency; $\earm$ quantifies the internal 
loss of the arm cavity (e.g., $\earm = 10^{-6}$ for 1ppm loss); $\esrc$ quantifies
the optical loss inside the SRC, 
including additional intra-cavity filters if any; 
$\eext$ denotes the external loss which includes the 
loss in the output chain and also the quantum inefficiency of the photodetection;
$\gamma$ is the bandwidth of the arm cavity and is equal to $c\,\Titm/(4 L)$
with $\Titm$ being the power transmission of ITM; 
$\alpha$ is equal to 1 if we use the internal squeezing to maximise
the power fluctuation 
and $\alpha$ is equal to $1/4$ if instead the internal squeezing 
is negligible, which has to do with the effect of internal squeezing on the signal 
response\,\cite{Peano2015, Korobko2017}, cf. Eq.\,\eqref{eq:sqz_sig} and also
Appendix\,\ref{app:OS-sqz};
 $\Tsrc$ is the effective transmissivity of SRC, which 
may be frequency dependent.  

\begin{figure}[!t]
  \includegraphics[width=\columnwidth]{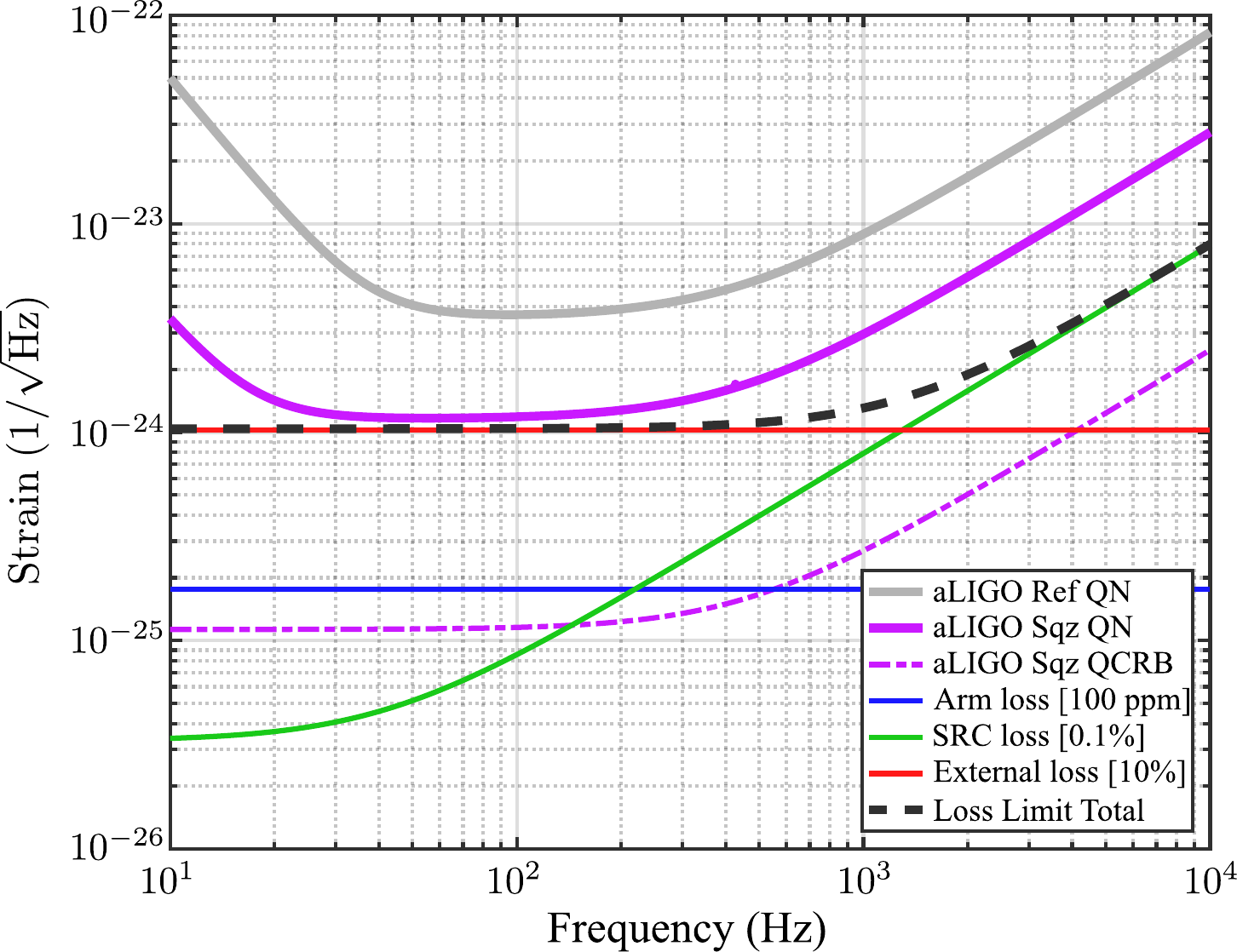}
    \caption{(Color online) This plot shows the quantum-noise 
(QN) sensitivity limits resulting from optical loss in an interferometer
 as given in Eq.\,\eqref{eq:Shh_loss} (blue, green, red and dashed-black curves).
For illustration, we assume a set of parameters similar to 
those of Advanced LIGO (aLIGO), and show the aLIGO QN for reference
 (upper light grey curve). 
To demonstrate the limited nature of the Quantum Cram\'er-Rao Bound (QCRB),
 the QN of aLIGO with a highly-squeezed input state is shown (solid purple curve)
 along with the corresponding QCRB (dot-dashed purple curve).
Notice how the QN of the highly-squeezed interferometer is limited by optical
 losses, while the QCRB is significantly less constraining.
Since the QCRB does not take into account losses,
 it can be made arbitrarily low by further increasing the squeezing of the input state, 
 i.e., reducing the uncertainty in the phase quadrature, and thereby increasing $S_{\!PP}$
 in Eq. \eqref{eq:FQL} beyond the 30\,dB level used for this example.
Note also that the highly-squeezed interferometer QN is above the loss limit (dashed-black curve)
 at low frequencies ($\lesssim$20\,Hz) due to the fixed read-out quadrature,
 and at high frequencies ($\gtrsim$400\,Hz) due to uncompensated dispersion.}
\label{fig:sensitivity}
\end{figure}

The arm cavity loss sets a flat limit across 
different frequencies, as the additional 
vacuum fluctuation is directly mixed with the signal inside the arm. 
The SRC loss is suppressed by ITM transmission at low frequencies; 
However, due to a finite arm cavity bandwidth which reduces the signal 
response, this effect becomes important at high frequencies.
The effect of the external loss 
depends on the transmission of SRC, which can be frequency dependent. 
For Advanced LIGO, $\earm$ is of the order of $10^{-4}$
($100\,\rm ppm$), $\esrc$ is around $10^{-3}$ which is mainly 
contributed from the beam splitter, and 
$\eext$ is around $0.1$ (coming from the mode-mismatch 
to the output mode cleaner and photodiode quantum inefficiency\,\cite{Evans:2013bs}). 
Given $\Titm= 0.014$ and 
$\Tsrc\approx 0.14$
in the default broadband detection mode, we show the resulting 
sensitivity bound imposed by current loss values in Fig.\,\ref{fig:sensitivity}. 

We can map this result to the one obtained in the quantum metrology 
community\,\cite{Dorner2009, Kolodynski2010, 
Knysh2011, Escher2011, Demkowicz-Dobrzanski2012} with a few simplifications.
The quantum metrology result considers
the effect of optical loss on the optimal phase estimation in Michelson or 
equivalent Mach-Zehnder interferometers. 
We can match this if we ignore the internal 
squeezing or focus at high frequencies where the internal optomechanical 
squeezing is weak, such that $\alpha=1/4$,
 and also ignore the SRC loss. The corresponding sensitivity limit 
from the arm cavity loss and the external loss is reduced to 
\begin{equation}
 S_{hh}^{\epsilon} = \frac{\hbar\, c^2} {4 L^2 \omega_0 P} \left( \epsilon_{\rm arm} + \frac{ T_{\rm src}}{4} \epsilon_{\rm ext}\right)\,.
\end{equation}
The extra factor of $T_{\rm src}/4$ in front of $\epsilon_{\rm ext}$ can be intuitively understood as the suppression of the loss effect from the signal recycling. 
 Furthermore, the impact of the losses discussed here are regularly included in the
 numerical computation of quantum noise in GW detectors\,\cite{BarsottiROP, Miao:2017uq}.
These numerical computations clearly show that losses limit the detector sensitivity,
 but they do not provide insight into the configuration-independent limit these losses impose.
Eq.\,\eqref{eq:Shh_loss} goes beyond the case-by-case approach followed thus far
 to provide a general loss limit for GW detectors. 

{\it Derivation ---} Here we show the details behind the main result 
Eq.\,\eqref{eq:Shh_loss}. We use the two-photon formalism developed by 
Caves and Schumaker\,\cite{Caves85}. In particular, we adopt the approach 
by Kimble {\it et al.}\,\cite{Kimble01} that is based upon this formalism 
and especially tailored to the context of laser interferometric GW detectors. 
In this formalism, the optical field at different locations 
is fully described by its amplitude quadrature $\hat a_1$
 and phase quadrature $\hat a_2$: 
\begin{equation}
\hat E(t) =\hat a_1(t)\cos\omega_0 t + \hat a_2(t)\sin\omega_0 t\,. 
\end{equation}
Moving into the 
frequency domain, $\hat a_{1,2}(\Omega)$ are labeled by the 
frequency $\Omega$, which is also called the sideband frequency and
 coincides with the GW signal frequency. 
Because the system is linear and time-invariant, 
different frequency components are independent of each other. At each frequency, 
the effects of different 
optical elements on the quadratures can be quantified by $2\times 2$
matrices, which act on the vector $\hat {\bm a}=(\hat a_1, \,\hat a_2)'$.
In particular, a passive element without external energy input
is described by a rotation matrix: 
\begin{equation}\label{eq:Mrot}
\Mrot = \left[
\begin{array}{cc}
\cos\Theta & -\sin\Theta \\
\sin\Theta & \cos\Theta \end{array}\right]\,.
\end{equation}
For example, the Fabry-Perot filter 
cavity used for frequency-dependent squeezing (or readout)
leads to a frequency-dependent rotation angle 
$\Theta(\Omega)$ for the amplitude and phase quadratures. 
The angle can be tuned by changing
 the cavity bandwidth and detuning\,\cite{Kimble01, Oelker2016}. 
In general, there will be an additional phase factor $e^{i\Phi(\Omega)}$ 
in front of the rotation matrix for the full description of a passive element. 

For a phase-sensitive active (squeezing) element\,\footnote{A phase insensitive
element will introduce an idler noisy channel\,\cite{Caves1982, Clerk2008}. The resulting 
sensitivity limit is in general larger than the phase sensitive one 
unless the idler channel is also measured and conditional upon. }, the corresponding 
matrix is 
\begin{equation}\label{eq:}
%{\bf M}_{\rm sqz}=\left[
%\begin{array}{cc}
% \cosh r-\cos 2\theta 
%   \sinh r & -\sin 2\theta
%   \sinh r \\
%- \sin 2\theta  \sinh r &
%   \cosh r+\cos 2\theta 
%   \sinh r \\
%\end{array}\right]\,.
\Msqz = \Mrot(\theta) \left[
\begin{array}{cc}
 {\rm e}^{r} & 0 \\
0 & {\rm e}^{-r} 
\end{array}\right] \Mrot(-\theta)\,.
\end{equation}
where $r$ is the squeezing factor and $\theta$ is 
the squeezing angle: $r = 1$ and $\theta = 0$ correspond 
to $\sim$9\,dB of phase 
squeezing.  In the most general cases, the
relevant parameters are frequency dependent,
namely $\theta=\theta(\Omega)$
and $r=r(\Omega)$. 
For example, the internal ponderomotive (optomechanical)
squeezing from the test-mass-light interaction inside the arm 
cavity can be described by the following matrix:
\begin{equation}\label{eq:Mpond}
\Mopt = \left[
\begin{array}{cc}
1 &0 \\
-\kappa & 1 \end{array}\right]\,,
\end{equation}
where $\kappa = 16 P\omega_0/(M c^2 \Omega^2)$.
It can be decomposed into the rotation matrix $\Mrot(\phi)$
followed by the general squeezing matrix $\Msqz(r, \theta)$, with
\begin{equation*}
\phi = -\arctan(\kappa/2 ), \,~\theta = {\rm arccot}(\kappa / 2)/2, \,~r = -{\rm arcsinh}(\kappa / 2)\,,
\end{equation*}
as shown explicitly in Refs.\,\cite{Kimble01, Kwee:2014cd}.

\begin{figure}[!t]
  \includegraphics[width=\columnwidth]{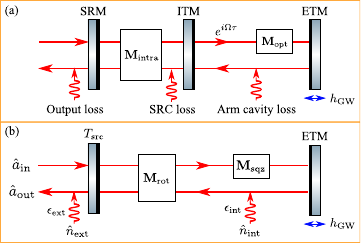}
    \caption{(a) A simplified schematic of the general
    scheme shown in Fig.\,\ref{fig:config} when only 
    focusing on the differential mode that contains the GW signal. The
    optical losses at different places are illustrated. (b) A further simplified 
    diagram by mapping the SRC
     into an effective mirror, and optical losses into internal and external components.
     Here $\Mrot$ accounts for a general rotation of the amplitude and 
     phase quadratures; $\Msqz$ describes the general squeezing effect.}
\label{fig:cavity}
\end{figure}

In Fig.\,\ref{fig:cavity} (a), we show a 
simplified representation of the general scheme in Fig.\,\ref{fig:config} when 
only looking at the differential mode, which contains the GW signal,
and the corresponding input and output fields at the differential (dark) port. 
We include the optical loss inside the arm cavity, 
the SRC and at the output. Depending on the 
configuration of the intra-cavity filters, the SRC loss can be contributed by 
several lossy channels. Each introduces a frequency-dependent loss
$\epsilon_i(\Omega)$. To derive the lower bound on the sensitivity 
without specifying the detail intra-cavity filter configuration, 
we assume a single effective frequency independent loss: 
\begin{equation}\label{eq:epsilonsrc}
\esrc = \min_{\Omega}\sum_i \epsilon_i(\Omega)\,, 
\end{equation} 
which is the minimum of the total loss over the frequency.   

Inside the arm cavity, the field not only experiences the ponderomotive  
squeezing mentioned earlier but also picks up a round trip 
phase of $2\Omega L/c$ through free propagation. 
The higher the frequency of the sideband, the more phase it picks up;
 this is sometimes referred to as ``positive dispersion''.
Such a phase relation makes the high-frequency signal sidebands 
deviate from the perfect resonant condition, which reduces 
their amplitude at the output.
In the case of Advanced LIGO, 
it decreases the detector sensitivity at high frequencies and 
leads to a finite detector bandwidth around 1kHz, as shown by 
the sensitivity curve in Fig.\,\ref{fig:sensitivity}. With the nominal configuration, 
one can increase the bandwidth by tuning the properties of the SRC. This comes, however, 
at the cost of decreasing the peak sensitivity. Such a trade-off between the bandwidth 
and peak sensitivity was first found by Mizuno\,\cite{Mizuno1995} and is a direct consequence of the QCRB in Eq.\,\eqref{eq:FQL}.

Adding any passive 
element to the intra-cavity filter will also introduce a frequency-dependent 
phase,
% $\Phi$ associated with the rotation,
 as mentioned after Eq.\,\eqref{eq:Mrot}.
This  trade-off cannot be circumvented with passive elements, but since we are only considering the sensitivity bound resulting from optical losses, we assume that the 
intra-cavity filters may also contain active elements that produce the so-called white-light-cavity effect.
With active elements present all signal sidebands may be
resonantly enhanced\,\cite{Wicht1997, Miao2015a, Zhou2015}. That is, these active elements produce the optimal phase $\phi_{\rm nd}$ with a negative dispersion, which cancels both the propagation phase $2\Omega L/c$ and the passive filter phase $\Phi$ around the frequency 
of interest, namely, 
\begin{equation}\label{eq:cancellation}
\phi_{\rm nd}(\Omega) = -\Phi(\Omega)-2\Omega L/c \,. 
\end{equation}
Any non-perfect cancelation will lead to a sensitivity that is worse than
the loss-induced limit considered here. 

With the assumption expressed in Eq.\,\eqref{eq:epsilonsrc} and
the condition given in Eq.\,\eqref{eq:cancellation}, we can further 
simplify the general scheme into what is shown in 
Fig.\,\ref{fig:cavity} (b): $\Mrot$ and $\Msqz$ 
capture the effect of both the internal squeezing  and 
the intra-cavity filters which gives rise to 
${\bf M}_{\rm intra}$ (a general rotation and squeezing matrix). 
The optical losses $\earm$ and $\esrc$ can then be combined to
 form a lower bound on the effective internal loss 
\begin{equation}
\eint = \earm + \frac{\Titm}{4} \left(1+\frac{\Omega^2}{\gamma^2}\right)\esrc\,. 
\end{equation}
The frequency dependent factor in front of $\esrc$ comes 
from the finite arm cavity bandwidth. We have assumed frequencies that 
are lower than the free spectral range (FSR) $f_{\rm fsr} = c/(2L)$ of the arm cavity. If this assumption is violated and the frequency is around other FSR, $\Omega$ must be replaced by $\Omega- 2 n \pi f_{\rm fsr}$ where $n$ is the index of the closest FSR $\Omega$. 

Following the propagation of the optical field and using the continuity condition at 
different interfaces, we can derive the frequency-domain input-output relation for 
the scheme in Fig.\,\ref{fig:cavity} (b):
\begin{equation}\label{eq:IO}
\hat {\bm a}_{\rm out} =
 \Mio\,\hat {\bm a}_{\rm in}   + \sqrt{\Tsrc\,\eint}\,\Mc 
\hat {\bm n}_{\rm int}  + \sqrt{\eext}\,\hat {\bm n}_{\rm ext}  + 
{\bm v}\, h_{\rm GW}\,, 
\end{equation} 
which quantifies the noise and signal content in the output at each frequency.
Here $\hat {\bm a}$ and $\hat {\bm n}$ with different subscripts 
are vectors of the amplitude and 
phase quadratures of fields at different locations. We do not include the 
input loss explicitly, which can be accounted for by assuming some degradation on 
the input squeezing level if the 
squeezed light is used. The transfer matrix $\Mio$ 
between $\hat {\bm a}_{\rm in} $ and $\hat {\bm a}_{\rm out} $ is defined as 
\begin{equation}
\Mio \equiv -\sqrt{R_{\rm src}} \,
{\bf I} +\Tsrc \Mc  \Mrot \Msqz \Mrot\,,
\end{equation}  
in which $\bf I$ is the identity matrix,
 $R_{\rm src}\equiv 1-\Tsrc$, and the  
transfer matrix  ${\bf M}_{\rm c}$ is defined as $[{\bf I}- \sqrt{R_{\rm src}}\, \Mrot \Msqz \Mrot]^{-1}$.
The vector  $\bm v$ describes the detector response to the GW signal: 
\begin{equation}\label{eq:vecv}
{\bm v}\equiv  \sqrt{\Tsrc}\, {\bf M}_{\rm c}\,\bm v_0
\end{equation}
with 
$\bm v_0=(0,\, \beta)'$ and $\beta=2\sqrt{\omega_0 L^2P/(\hbar c^2)}$.

Using the homodyne readout, we can measure the general quadrature of 
the outgoing field $\hat a_{\rm out}(\zeta)$ which is equal to 
$(\cos\zeta,\,\sin\zeta)\cdot \hat {\bm a}_{\rm out}$. 
If we apply an output filter that optimises $\zeta$ at 
all frequencies, we will obtain the best sensitivity with 
the minimum signal-referred noise spectral density:
\begin{equation}\label{eq:Shmin}
S^{\rm min}_{hh}=\frac{1}{{\bm v}^{\dag} {\bm \Sigma}^{-1}_{\rm tot} {\bm v}}
\approx S_{hh}^{\rm QCRB} + S_{hh}^{\epsilon}\,,
\end{equation}
where ${\bm \Sigma}_{\rm tot}\equiv
 {\bf M}_{\rm io}{\bf M}_{\rm io}^{\dag}+\Tsrc\eint 
  {\bf M}_{\rm c}{\bf M}_{\rm c}^{\dag}+
 \eext {\bf I}$ is the total covariance matrix, and the 
 approximation is to the first order in
 $\eint$ and $\eext$.

Since the interferometer under consideration is a linear Gaussian system, we can 
view the signal part of Eq.\,\eqref{eq:vecv} as displacing the mean of 
the Gaussian state of the optical field. We can obtain the sensitivity
bound by evaluating the quantum Fisher information for estimating 
the mean of a Gaussian state\,\cite{Pinel2012, Monras2013, Gao2014, Safranek2015, Branford2018}.
This approach gives exactly the same result shown above, i.e. 
Eq.\,\eqref{eq:Shmin} is also the QCRB in the presence of optical loss and 
the optimal homodyne detection is the one that saturates it. We intentionally 
separate the first term and call it as $S_{hh}^{\rm QCRB}$ to echo the result presented in Ref.\,\cite{Braginsky2000EQL, Miao2017} where the lossless case was considered. 
 
The explicit form of $S^{\rm QCRB}_{hh}$ and $S_{hh}^{\epsilon}$, 
in terms of $r$, $\Tsrc$, 
$\Theta$, and $\theta$ is quite complicated. However, not all parameter regimes
are relevant. To achieve a low QCRB or large power fluctuation, 
we require $\Tsrc\ll 1$ to enhance the signal recycling. We also assume
$\Theta\ll 1$ to focus on relevant frequencies that are 
within one free spectral range of the arm cavity. 
The internal squeezing needs to be 
of the same order of $\Tsrc$ so that 
the round-trip gain of the amplitude quadrature is close to unity, which 
can then result in a significant level of power fluctuation. In such 
a parameter regime, we can make 
a Taylor expansion of both $S^{\rm QCRB}_{hh}$ and $S_{hh}^{\epsilon}$ 
with respect to these small parameters. Specifically, 
up to the leading order of $\Tsrc$, $\Theta$, and $r$, we have 
\begin{equation}
S^{\rm QCRB}_{hh} =\frac{\hbar\,  c^2 (\delta^2-4 r^2 )^2e^{-2r_{\rm input}}}{16L^2 \omega_0
   P \Tsrc[\delta^2+4r^2+4 \delta r \sin(\theta+\theta_0)]}\,, 
\end{equation}
where $r_{\rm input}$ is the squeezing factor of the 
input squeezed light, $\delta\equiv \sqrt{\Tsrc^2+16 \Theta^2}$ and $\theta_0\equiv 
\cot^{-1} (4\Theta/\Tsrc)$.
This means that we can make the above QCRB vanish, i.e., achieving an unbounded 
sensitivity in the ideal lossless case, if $r=\delta/2$.
Under this condition, the resulting 
sensitivity limit due to the optical loss reads
\begin{equation}\label{eq:shh_loss1}
\min_{\theta} S^{\epsilon}_{hh}= \frac{\hbar\,  c^2 }{4 L^2 \omega_0
   P}(\eint+\Tsrc\, 
    \eext)\,,
\end{equation}
where the minimum is achieved by setting $\theta=\pi/2+\theta_0$. 

If however the internal squeezing is negligible with $r=0$, 
the QCRB cannot be made 
to be arbitrary small given a finite input squeezing, 
and we have 
\begin{equation}
S^{\rm QCRB}_{hh}=
\frac{\hbar\,  c^2 \delta^2 e^{-2r_{\rm input}}}{16 \Tsrc L^2 \omega_0
   P}\,,
\end{equation}
which is simply the shot-noise-only sensitivity given a general $\Theta$. 
The corresponding 
 loss-induced  limit is
\begin{equation}\label{eq:shh_loss2}
\min_{\phi}S^{\epsilon}_{hh}=\frac{\hbar\,  c^2 }{4 L^2 \omega_0
   P}\left(\eint+\frac{\Tsrc}{4}\, 
    \eext\right)\,,
\end{equation}
where the minimum is attained when $\Theta=0$ (tuned). 
Eqs.\,\eqref{eq:shh_loss1} and \eqref{eq:shh_loss2} together give 
our main result shown in Eq.\,\eqref{eq:Shh_loss}, when expanding out
$\eint$ in terms of $\earm$ and $\esrc$. 

Worthy of pointing out, there is a factor of four difference 
in the dependence of $\eext$ 
between Eq.\,\eqref{eq:shh_loss1} and Eq.\,\eqref{eq:shh_loss2}. 
It originates from a reduced signal response when 
trying to maximise the sensitivity using the internal squeezing: 
\begin{equation}\label{eq:sqz_sig}
\frac{|{\bm q}_{\zeta_{\rm opt}} {\bm v}(r=\delta/2) |}
{|{\bm q}_{\zeta}\,{\bm v}(r=0)|}
=\left[\frac{1+\sin(\theta+\theta_0)}{4}\right]^{1/2}\le \frac{1}{2}\,,
\end{equation}
which shows that the signal response with internal squeezing 
is at least factor of two smaller. 
Such a reduction of signal response
was mentioned by Peano {\it et al.}~\cite{Peano2015} and
recently investigated experimentally by Korobko {\it et al.}~\cite{Korobko2017}, 
when considering placing an 
internal frequency-independent squeezer inside an optical cavity. 
In the case of laser interferometric GW detectors, even without introducing 
an additional squeezer, as mentioned earlier, there is the internal ponderomotive 
squeezing. The only difference is that the corresponding squeezing factor $r$ is highly 
frequency dependent due to the test mass response, cf. Eq.\,\eqref{eq:Mpond}, and the condition $
r=\delta/2 $ can only be satisfied at a single frequency, given the nominal 
dual-recycled Michelson interferometer, which is also illustrated in Appendix\,\ref{app:OS-sqz}.
 
{\it Conclusions and Discussions ---} We present a new fundamental 
limit to gravitational-wave detector sensitivity based on optical losses 
which lead to decoherence.
While the quantum Cram\'er-Rao bound in Eq.\,\eqref{eq:FQL} 
provides a fundamental limit
to the sensitivity of gravitational-wave detectors in the 
ideal lossless case, the optical-loss-induced limit presented
 in Eq.\,\eqref{eq:Shh_loss} can be more stringent.
Unlike the QCRB, the loss limit cannot be made irrelevant
with high levels of external or internal squeezing. The implication of our 
study for future gravitational-wave 
detectors is that the minimization of
optical losses in the interferometer arms and in the 
readout must be a strong focus of research and development efforts. 

There are three additional points that we would like to mention to broaden 
the scope and applicability of our result.
Firstly, while we have focused our discussion on Michelson-type interferometers
there are advanced QND configurations based on Sagnac 
interferometers\,\cite{Chen2003, Graf2014}.
Optical loss is also an important limiting factor for such 
quantum speed-meters\,\cite{Danilishin2015} and our result can be directly applied to the 
equivalent sloshing-cavity-based speed-meter scheme\,\cite{Purdue2}. For 
Sagnac speed-meters with additional intra-cavity and output filters, the term in Eq.\,\eqref{eq:Shh_loss} from the arm-cavity loss is still the same, but we need to 
introduce an additional factor of $(\omega_s/\Omega)^2$ at frequencies below 
$\omega_s$ (the characteristic frequency of the speed response) for the limit from 
the SRC loss and the external loss. This factor accounts for the difference between the position response and the speed response at low frequencies. 

Secondly, our result also applies to any optomechanical sensors that can be put into
the general scheme illustrated in Fig.\,\ref{fig:cavity} (b). They share
the same principle as the laser interferometric gravitational-wave detectors, even though they may operate in a different parameter 
regimes and have different forms of ponderomotive squeezing\,\cite{Chen2013, Aspelmeyer2014}. The only issue occurs around the resonant frequency of the mechanical oscillator where, if the optical loss is small, the loss limit may not be the dominant one and the zero-point fluctuation (ZFP) of the mechanical oscillator will impose a more stringent bound\,\cite{Caves1982, Clerk2008}. Moreover, the quantum noise of the light cannot be made lower than the mechanical zero-point fluctuation at the resonance using the homodyne detection\,\cite{Buchmann2016, Kampel2017, Mason2018}; the total quantum noise is actually twice (in power) the noise from zero-point fluctuations---an interesting consequence of the linear quantum measurement theory\,\cite{BK92, Khalili2012}. 

Finally, as illustrated in Appendix\,\ref{app:OpFDT}, the loss effect described
here is analogous to thermal noise from the mechanical dissipation,
which provides another fundamental limit to gravitational-wave detectors and other high-precision measurements. 
This not only highlights the fundamental role of optical losses in quantum metrology, 
but also motivates future studies which may eventually allow us to understand
both the mechanical dissipation  and the optical dissipation under a unified framework.

{\it Acknowledgements ---} We would like to thank Rana Adhikari, 
Lisa Barsotti, Yanbei Chen, Stefan Danilishin, Jan Harms,
Farid Khalili, Mikhail Korobko, Yiqiu Ma, Denis Martynov, 
members of the LSC AIC, and QN groups for 
fruitful discussions. H.M. is supported by UK STFC 
Ernest Rutherford Fellowship (Grant
No. ST/M005844/11). M.E. acknowledges 
the support of the National Science Foundation 
and the LIGO Laboratory. 
LIGO was constructed by the California Institute
 of Technology and Massachusetts Institute of Technology with funding
 from the National Science Foundation and operates under cooperative
 agreement PHY-0757058.

\appendix

\section{Understanding the optical spring from internal ponderomotive squeezing}
\label{app:OS-sqz}

Here we provide some insights into the optical spring effect in the laser 
interferometric GW detectors from the perspective of 
internal ponderomotive squeezing. Conventionally, the optical spring is understood 
in terms of a modification to the dynamics of the test mass through an optical feedback; the free mass is effectively turned into a harmonic oscillator that resonates at the 
optical spring frequency\,\cite{BuCh2002}. In this picture, 
the enhancement of the sensitivity is 
attributed to an increase of the response to the GW signal (viewed as a tidal force
acting on the test mass). However, as we have seen from the discussion about the 
internal squeezing, if we choose the optimal readout quadrature for 
maximising the sensitivity, cf. Eq.\,\eqref{eq:sqz_sig}, the signal response is decreased by a factor of two compared with the case without using the internal squeezing. 
This seems to be in contradiction to the interpretation of the optical spring effect as arising from the internal ponderomotive squeezing\,\cite{Miao2017}. It can be clarified if we look at the  the noise amplitude and signal response that define the sensitivity separately. 

\begin{figure}[!t]
  \includegraphics[width=\columnwidth]{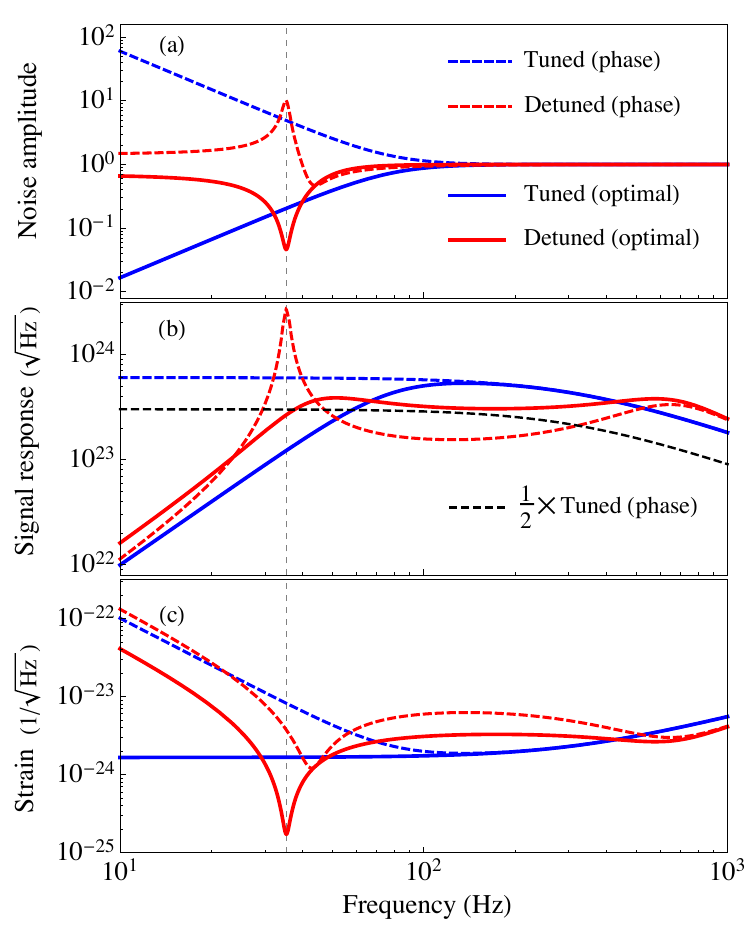}
   \caption{(a) The noise amplitude given the measurement of output phase quadrature
    (dashed line) and the optimal quadrature (solid line) in the two cases: tuned SRC (blue) 
    and detuned SRC (red), respectively. (b) The corresponding signal response for these two cases. We also show the half of the signal response with a phase measurement 
    in the tuned case as a reference (dashed black curve). (c) The sensitivity curve which is a ratio of the noise amplitude in (a) and the signal response in (b). We assume the interferometer parameters the same as the aLIGO design but choosing the detector bandwidth to be 240 Hz and detuning to be 600 Hz for illustrations. The optical spring
    frequency is marked using the vertical grid line around 35 Hz. }
\label{fig:OS_sqz}
\end{figure}

In Fig.\,\ref{fig:OS_sqz}, we compare two cases: one with tuned SRC and the other with detuned SRC---the optical spring effect is present in the 
latter. At the optical spring frequency, the signal response is amplified compared to the tuned case without the optical spring, when we measure the output phase quadrature (the default quadrature that GW detectors measure). However, the 
corresponding noise amplitude is also high, leading to a suboptimal sensitivity. If instead  the optimal quadrature is measured, we can achieve the best sensitivity at the optical 
spring frequency, which was first presented in Ref.\,\cite{Harms2003}. The main contribution comes from a significant reduction of the 
noise amplitude due to the internal ponderomotive squeezing. The signal response for the optimal quadrature readout is indeed reduced by a factor of two compared with the tuned case with phase quadrature measurement, agreeing with our statement earlier. Therefore, reaching the optimal sensitivity using the optical spring effect is
mostly attributable to the squeezing rather than the signal enhancement. Worthy of mentioning, we are considering the presence of a small amount of optical loss. If the output loss is very high and significantly degrade the generated
squeezing, the optimal quadrature for maximising the
sensitivity can deviate from the one with a reduced signal response, and having signal enhancement can be preferrable\,\footnote{This point was raised by Mikhail Korobko in a private communication.}. 

\section{Optical fluctuation-dissipation theorem}
\label{app:OpFDT}

The loss-induced quantum limit can be viewed as arising from 
the optical dissipation, analogous to the thermal noise from the mechanical 
dissipation. Such an analogy, in some limiting cases, can 
be made exact by using
the fluctuation-dissipation theorem (FDT)\,\cite{Kubo1966}. In GW community, 
the FDT has been widely applied for 
analysing other classical fundamental limits to GW detectors, arising
 from the thermal noise in
the suspension and test-mass mirrors\,\cite{Saulson:book, Levin:Direct, 
Harry:AO2006}. The idea of applying FDT to the optical degrees 
of freedom was first envisioned by one of us Nicolas D.
Smith and Yuri Levin\,\footnote{Private communications.}, and the goal 
is to provide a unified treatment of the quantum noise and the classical
thermal noise. However, the non-equilibrium nature of 
the quantum measurement process defies a straightforward 
generalisation. In particular, we have a coherent laser field driving the 
system out of equilibrium which induces internal ponderomotive
squeezing. 
Interestingly, as we have seen from the previous analysis, the internal 
squeezing only changes the loss limit by a factor of 4 without changing
the dependence to other parameters. Therefore, the
equilibrium FDT applied to the case without the internal squeezing
can provide some quantitative and useful insights into the loss limit, 
as what we are trying to show in the discussion below. 

Since we are looking at the optical frequency, 
the external continuum field introduced by the optical loss, to a good approximation, can be treated as a zero-temperature heat bath.  
As a result, according to the FDT\,\cite{Callen1951,Kubo1966}, any
dynamical quantity $x$ of the system, which couples to the continuum field, 
satisfies 
\begin{equation}\label{eq:FDT}
S_{xx}(\omega)=2\hbar \, {\rm Im}[\chi_{xx}(\omega)]\,, 
\end{equation}
where $\chi_{xx}$ is the susceptibility of $x$ 
and ${\rm Im}[\chi_{xx}]$ 
means its imaginary part which quantifies the dissipation. 

In our case, the system is the optical field at different locations which couples 
to the heat bath due to presence of the optical loss and also responds 
to the GW signal. Let us use the field inside the arm cavity and the 
associated arm cavity loss to illustrate the basic idea. 
At frequencies below the free spectral range 
$c/{2L}$, the arm cavity field can be modelled as a single mode or 
a simple damped harmonic oscillator: its amplitude quadrature $\hat A_1$ 
and phase quadrature $\hat A_2$ satisfy the following equations of motion:
\begin{align}
\dot {\hat A}_1(t)+\gamma_{\epsilon}\hat A_1(t) &
= \omega_{\rm cav} \,\hat A_2(t)+\sqrt{2\gamma_{\epsilon}}\, \hat A_1^{\rm ext}(t)\,,\\
\dot {\hat A}_2(t)+\gamma_{\epsilon}\hat A_2(t) &= -\omega_{\rm cav} \,\hat A_1(t)
+\sqrt{2\gamma_{\epsilon}}\,\hat A_2^{\rm ext}(t)\,,
\end{align}
where $\omega_{\rm cav}$ is the cavity resonant frequency,  
$\gamma_{\epsilon}=c \, \earm/(4L)$ is the damping 
rate due to the arm cavity loss, and $\hat A^{\rm ext}$ is the external continuum 
field. In the linear-response theory\,\cite{Kubo1966, BuCh2002, Miao2017}, the susceptibility is defined as 
\begin{equation}
\chi_{AB}(t-t')\equiv (i/\hbar)[\hat A(t),\, \hat B(t')]\Theta(t-t')\,.
\end{equation}
Using the commutator relation $[\hat A_1^{\rm ext}(t),\, \hat A_2^{\rm ext}(t')]= i \hbar \delta(t-t')$, 
we can derive the relevant susceptibilities $\chi_{A_2 A_2}$ and $\chi_{A_2 A_1}$. 
In the frequency domain, they can be written as  
\begin{align}\label{eq:chi_22}
\chi_{A_2 A_2}(\omega)& =\frac{\omega_{\rm cav}}
{\hbar[(\gamma_{\epsilon}-i\omega)^2+\omega_{\rm cav}^2]}
\,, \\\label{eq:chi_21}
\chi_{A_2 A_1}(\omega)& =\frac{i\omega-\gamma_{\epsilon}
}{\hbar[(\gamma_{\epsilon}-i\omega)^2+\omega_{\rm cav}^2]}\,.
\end{align}

In the transverse-traceless (TT) gauge~\cite{Cooperstock1993, Tarabrin2008}, 
a GW acts as a strain directly coupled to the cavity mode, of which 
the linearized interaction
Hamiltonian is 
\begin{equation}\label{eq:Hint}
\hat H^{\rm TT} = -\hbar g \hat A_1\, L \, h_{\rm GW}\,.
\end{equation}
Here we have defined $g\equiv 2\sqrt{P\omega_0/
(\hbar L c )}$ and $h_{\rm GW}$ is the GW strain. 
With the FDT in Eq.\,\eqref{eq:FDT} and 
the susceptibility Eq.\,\eqref{eq:chi_22}, 
the sensitivity limit due to arm cavity loss can be 
obtained by normalising the fluctuation
of the phase quadrature with respect to the signal response 
obtained 
from Eqs.\,\eqref{eq:chi_21} and \eqref{eq:Hint}, namely, 
\begin{equation}\label{eq:Shh_loss_int1}
S^{\epsilon}_{hh}(\omega)=\frac{S_{A_2 A_2}(\omega)}{\hbar^2 g^2 L^2 |\chi_{A_2A_1}(\omega)|^2}=
\frac{ 2\,{\rm Im}[\chi_{A_2 A_2}(\omega)]}{\hbar g^2 L^2 |\chi_{A_2A_1}(\omega)|^2}
\end{equation}
Switching to the sideband frequency $\Omega$ with  
the laser frequency $\omega_0$ as the reference, we obtain the 
first term in Eq.\,\eqref{eq:Shh_loss}:
\begin{equation}\label{eq:Shh_loss_int2}
S^{\epsilon}_{hh}(\Omega)=\frac{4\gamma_{\epsilon}\omega\,\omega_{\rm cav}}
{(\omega^2+\gamma_{\epsilon}^2)g^2 L^2}\approx \frac{\hbar c^2 \earm}
{4 L^2 \omega_0 P}\,,
\end{equation}
where we have used the fact that $\omega_{\rm cav}$ is approximately equal to 
$\omega$ with $\omega=\Omega+\omega_0\approx \omega_0$, and $\omega_0\gg \gamma_{\epsilon}$. 

The above result still holds even when the arm cavity mode coupled to 
additional degrees of freedom. 
This is because the ratio ${\rm Im}[\chi_{A_2 A_2}]/|\chi_{A_2 A_1}|^2$ 
is an invariant, as long as
there is no dissipation in these additional degrees of 
freedom and they are passive. 
To prove this, we can have the phase quadrature of the cavity 
mode coupled to
some general coordinate $\hat y$ of one of these 
degrees of freedom. Using the 
linear-response theory,
 the coupling
modifies the original susceptibilities $\chi_{A_2 A_2}$ 
and $\chi_{A_2 A_1}$ into 
\begin{equation}\label{eq:}
\chi^{\rm new}_{A_2 A_2}=\frac{\chi_{A_2A_2}}
{1-\chi_{A_2A_2}\chi_{yy}}\,,\quad \chi^{\rm new}_{A_2 A_1}= \frac{\chi_{A_2A_1}}
{1-\chi_{A_2A_2}\chi_{yy}}\,. 
\end{equation}
Since no dissipation is present in these degrees of freedom, i.e., 
${\rm Im}[\chi_{yy}]=0$, we have 
\begin{equation}\label{eq:ratio}
{\rm Im}[\chi^{\rm new}_{A_2 A_2}] = \frac{{\rm Im}[\chi_{A_2A_2}(1-\chi^*_{A_2A_2}\chi_{yy})]}
{|1-\chi_{A_2A_2}\chi_{yy}|^2}= \frac{{\rm Im}[\chi_{A_2A_2}]}
{|1-\chi_{A_2A_2}\chi_{yy}|^2}\,. 
\end{equation}
This implies ${\rm Im}[\chi^{\rm new}_{A_2 A_2}]/|\chi^{\rm new}_{A_2 A_1}|^2={\rm Im}[\chi_{A_2 A_2}]/|\chi_{A_2 A_1}|^2$, and regardless of the type of intra-cavity filters introduced, $S^{\epsilon}_{hh}(\Omega)$ is given 
by Eq.\,\eqref{eq:Shh_loss_int2} for arm cavity loss. 

Similarly, we can derive the result for the SRC
loss and the output loss by converting them into an effective arm cavity loss. However,
such a conversion can be made exact 
only when we ignore the internal squeezing or 
tweak the signal response by compensating the additional factor of four difference mentioned earlier. 

%%%%%% References %%%%%%

\bibliography{references}

%%%%%% End %%%%%%

\end{document}